\documentstyle[aps,multicol,epsf,amstex]{revtex}

\draft
\def\squote{}
\def\quote#1#2#3#4{\squote {#1,\ {\sl#2}\ {\bf#3}, #4}.\par} 
\def\qquote#1#2#3#4{\squote {#1,\ {\sl#2}\ {\bf#3}, #4};}

\def\prl{{\sl Phys. Rev. Lett.}\ }

\def\pr {{\sl Phys. Rev.}\ }
\def\ksi{\xi}

\def\e{\epsilon}

\begin{document}
\title{Quantum Hall Effect in Three Dimensional Layered Systems}
\bigskip
\author{\large Yigal Meir}
\address{Department of Physics, Ben Gurion University, Beer Sheva 84105, ISRAEL}
\maketitle
\begin{abstract}
Using a mapping of a layered three-dimensional system
  with significant inter-layer
tunneling onto a spin-Hamiltonian,  the phase diagram in the strong magnetic
field limit is obtained in the semi-classical approximation. This phase
diagram,  which exhibit a metallic phase for a finite range of energies and
magnetic fields,  and the calculated associated critical exponent,  $\nu=4/3$, 
agree excellently with existing numerical calculations. The implication of this
work for the quantum Hall effect in three dimensions is discussed.
\end{abstract}
\begin{multicols}{2}
The quantum Hall effect is one of the hallmarks of two-dimensional electron 
systems 
\cite{vonklitizing,stone}. The possibility of the occurrence of the quantum Hall
effect in three dimensions was explored rather early \cite{azbel},  and precursors
of the quantum Hall effect were observed in some three dimensional systems 
\cite{3dexperiments}. The existence of well quantized Hall plateaus 
was, however, demonstrated only in  three dimensional
layered semiconductors   with significant interlayer coupling \cite{stormer}.
 These layered systems have attracted significant
theoretical interest recently,  due to the proposed existence of a metallic
phase for a finite range of energies or magnetic fields \cite{chalker},  and a 
new ``chiral'' two-dimensional metallic phase on the surface \cite{chiral}.
The existence of such a metallic phase at the surface was recently confirmed
experimentally in measurements of the vertical conductance ($\sigma_{zz}$) 
\cite{beth}.

In this work we use a mapping of the three-dimensional layered structure 
onto a two-dimensional spin-Hamiltonian. Using a semi-classical description
we derive the phase-diagram \cite{chalker} and obtain the critical
exponent $\nu$,  describing the divergence of the localization length $\ksi$, 
as one approaches the transition from the insulating side, 
 $\ksi\sim |E-E_c|^{-\nu}$,  or $\ksi\sim |B-B_c|^{-\nu}$,  where $E_c$ and
 $B_c$ are the critical energy and magnetic field,  respectively. The derived
 critical exponent $\nu=4/3$ agrees excellently with existing numerical
 data,  $\nu=1.35\pm0.15$,  obtained both for a layered system and
  three-dimensional tight-binding model \cite{kramer},  and $\nu=1.45\pm0.25$, 
  obtained in Ref.\cite{chalker} from a layered network model \cite{network}. 
  
\begin{center}
\leavevmode
\epsfxsize=3.4in
\epsfbox{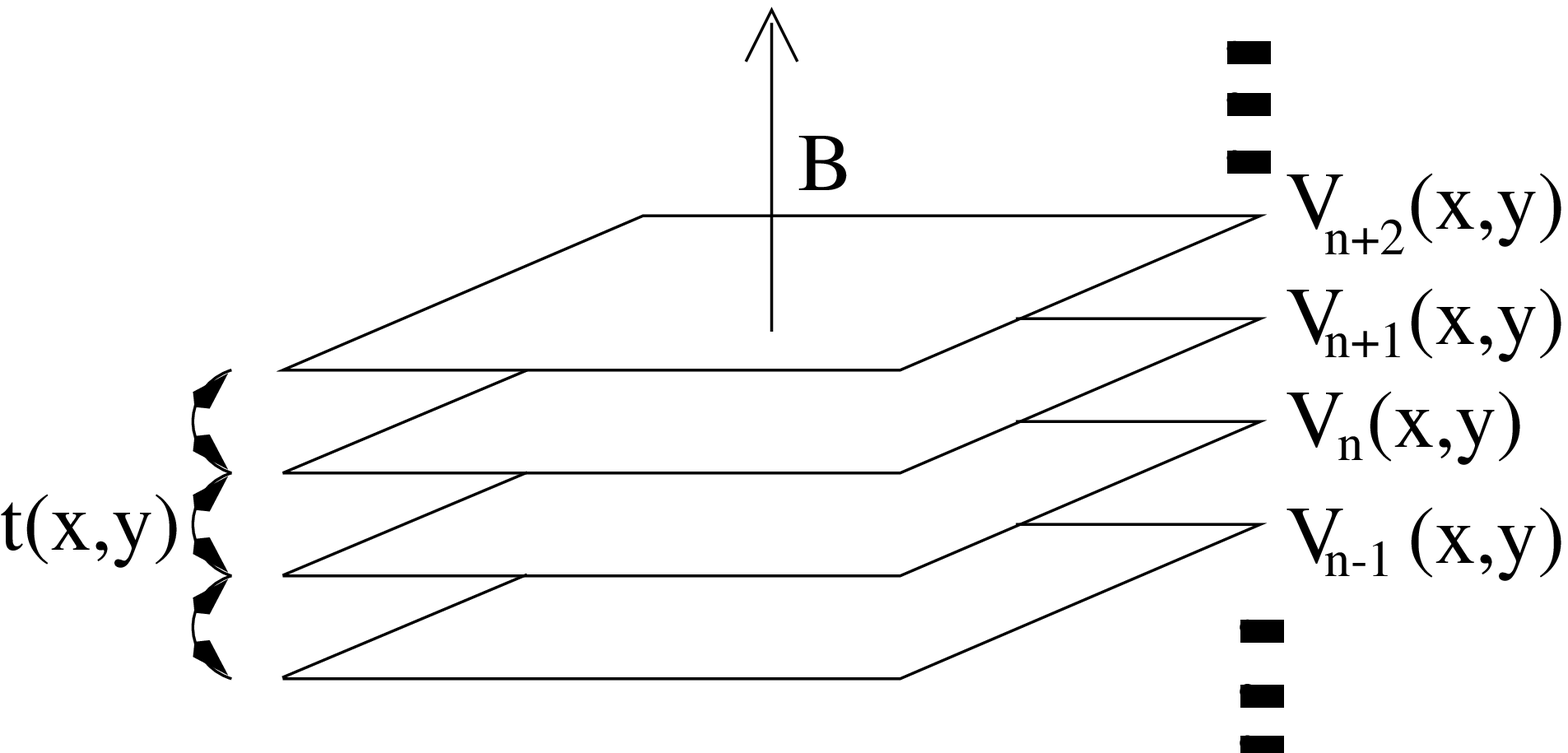}
\end{center}
\begin{small}
\vskip 1 truecm
\centerline{Fig. 1. The system studied in this work.}
\end{small}
\vskip 0.5 truecm

We start with the Hamiltonian describing a spinless electron in a system
of N coupled two-dimensional layers (see Fig. 1), 
\begin{equation}
{\cal H} = \sum_i^N \left[ ({\bf p}_i- e {\bf A}_i/c)^2/2m + V_i(x, y) + 
T_i(x, y)
\right] , 
\label{hamilonian}
\end{equation}
where $x$ and $y$ are coordinates in the plane and  the summation is over the
layers. The first and second terms in the brackets
 describe the kinetic and potential
energies within a layer,
  while the third term describe the hopping between adjacent
layers, which may depend on the position 
in the plane. The layer potentials are assumed to be independently distributed
with zero mean.

We now associate with the electron a spin-index that corresponds to the layer
index in (\ref{hamilonian}). The interlayer tunneling will now correspond 
to spin-raising and lowering operations. In order to describe the different
potential landscape in each layer,  we add  a random $S_z$
term to the Hamiltonian that now describes a spin-$S$ ($=(N-1)/2$) electron
moving in two dimensions, 
\begin{eqnarray}
{\cal H} &=& ({\bf p}- e {\bf A}/c)^2/2m + U(x, y) + \Delta U(x, y) S_Z
\label{spinhamilonian}
\\ \nonumber
 &+& t(x, y) S_+ + t^*(x, y) S_-  .
\end{eqnarray}
The second term describes a random potential independent of the spin (layer)
index. The third term accounts for the different potentials for the different
spin direction,  by a random shift of the potential between adjacent layers 
(at each point of the plane). Thus at each point the electron sees a different
potential in each layer (or for each spin direction). Since the shift 
$\Delta U(x, y)$ is random in sign and in magnitude,  the average potential in
each layer is the same \cite{but}.

The Hamiltonian (\ref{spinhamilonian}) can now be simply written as
\begin{equation}
{\cal H} = ({\bf p}- e {\bf A}/c)^2/2m + U(x, y) + 
{1\over S}{\bf S}\cdot {\bf H}(x, y) , 
\label{hhamiltonian}
\end{equation}
namely a spin-$S$ electron moving in two-dimensions under the influence
of a random potential and a random magnetic field (coupled to its spin).
The advantage of this representation is that one can try to generalize
methods that worked for the two-dimensional case,  in the absence of a
random field,  to include the effects of the field. In the following
we will concentrate on the large (uniform) magnetic field limit,  where
the kinetic energy is quenched and one may treat the electrons semi-classically.
In the absence of the random field the electron moves along equi-potential lines.
As is well known in this case \cite{ramit},  
electrons with too small an energy will be 
trapped around potential valleys,  while for too high an energy they will be
trapped around potential hills. There is a single ``critical'' energy  where
the electron trajectory percolates through the system. This corresponds to
 the quantum Hall transition,  where there is a single energy (at the center
 of the Landau level in case of symmetrically distributed random potentials)
 where states are extended.
 
 In the present case,  in the same strong magnetic field limit,  it is the
 total energy -- the potential energy plus the spin energy (due to the random
 field) that is conserved. Thus,  as the electron rotates its spin along the
 trajectory,  it exchanges energy between the potential energy and the
 spin-energy,  such that the total is conserved. The range of potential
 energies accessible by the electron has a width $\Delta\equiv2H_R$,  where 
 $H_R$ is the typical amplitude of the random field. Consequently,  even if the
 electron does not have the correct (critical) potential energy to percolate
 through the system to begin with,  it can still do that as long as its
 total energy is within $H_R$ of the critical energy.

\begin{center}
\leavevmode
\epsfxsize=3.3in
\epsfbox{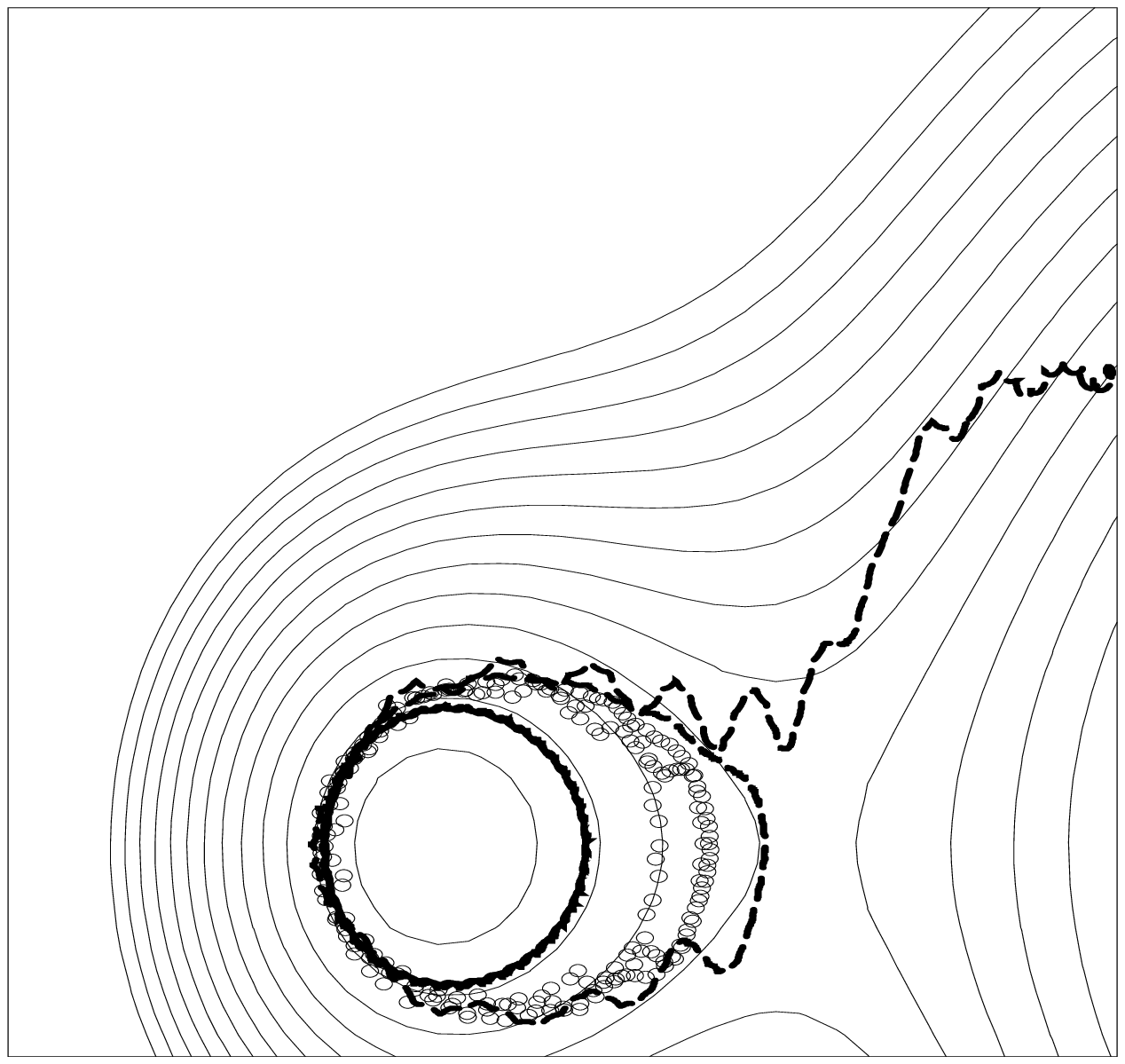}
\end{center}
\begin{small}
\vskip -1.5 truecm
Fig. 2. The classical trajectories of an electron in a strong magnetic
 field. Without random field,  the electron follows equipotential lines (solid
 curve); with increasing random field it explores larger portion of the potential
 energy landscape (circles),  until for large enough random field,  it can go
 through the saddle point (broken line).
\end{small}
\vskip 0.5 truecm

 An example is depicted in Fig. 2. The classical equations of motion for the
 Hamiltonian (\ref{hhamiltonian}) with $U(x,y)$ corresponding to two impurities
 (the equipotential lines appear as thin solid curves) were integrated. In the
 absence of a random field (a solid thick curve),
 the electron follows a single equipotential line,  with superimposed 
 cyclotron oscillations,  and is trapped around one impurity. With increasing
random field the electron explores a larger portion of the potential energy
landscape (see,  e.g., the trajectory denoted by circles),  until,  for
large enough random field (broken line),  the electron can go through
 the saddle point and percolate away. In the original layered system, this 
 process corresponds to the possibility of the electron tunneling to a different
 layer and drifting along a different potential line (with the same potential
 energy). Thus as the energy is increased, before percolation occurs in
 a single layer,  there will be a percolating path
 consisting of equipotential lines in different layers,
 connected by inter-layer tunneling events.

Since the random magnetic field amplitude $H_R\sim\sqrt{t^2+(\Delta U)^2}$, one
expects a region of extended states that increases with $t$,  leading to the 
phase diagram depicted in Fig. 3. For any finite $t$ there exists a finite
range of energies (or magnetic fields) where the system is metallic. Accordingly, 
even at $T=0$ the transition between Hall plateaus will not be sharp, 
 but rather
occur in a finite range of magnetic fields or gate voltages.

Interestingly,  in the present semi-classical 
description such a metallic phase will occur 
even for an infinitesimal tunneling matrix element $t$. The reason is that
once $t\ne0$ the electron can,  in principle,  rotates its spin (tunnel between
layers) and explore the whole energy range allowed by conservation of total
energy. We know,  however, that quantum mechanically, 
for small enough tunneling matrix element, the electron will be localized in
spin-space and the range of potential energies available (i.e. the width of
 the metallic region in phase space) will be much smaller than one expects
 classically,  going to zero as $t\rightarrow0$ \cite{exponent}. Thus,  there
 is a region in the phase diagram (the shaded part of Fig. 3),  where the
 electron is localized quantum mechanically,  but its classical trajectory
 is extended. The derived phase diagram (Fig. 3) agrees with the phase diagram 
 established numerically by Chalker and Dohmen \cite{chalker}.

\begin{center}
\leavevmode
\epsfxsize=3.3in
\epsfbox{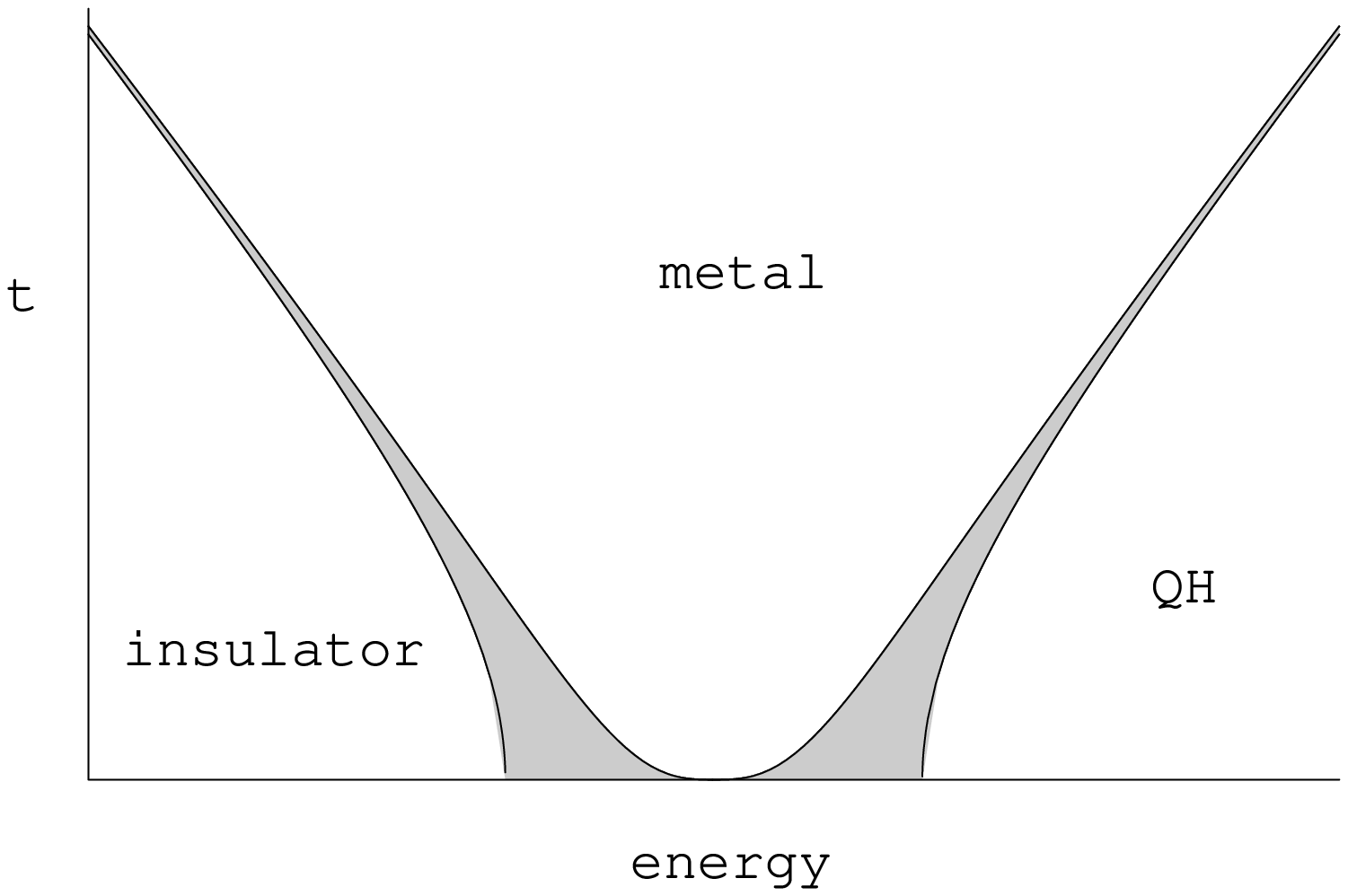}
\end{center}
\begin{small}
\vskip -2.5 truecm
Fig. 3. Phase diagram of the layered system. For finite inter-layer tunneling
$t$ there is a finite range of energies where extended states exist. The shaded
region is where the wavefunctions are localized,  but the classical trajectories
are extended (see text).
\end{small}
\vskip 0.5 truecm

We now turn to the critical behavior. For the two-dimensional
quantum Hall problem Mil'nikov and Sokolov 
\cite{sokolov,phase} used the following  argument to predict the 
critical exponent. In the classical description,  away from
the critical energy $E_c$, the electron is confined to a percolation cluster
of typical size $\ksi_p$,  the percolation coherence length. Near the threshold 
$\ksi_p\sim|E_c-E|^{-\nu_p}$, where $\nu_p=4/3$ is the two-dimensional percolation
exponent. As one approaches the transition the clusters approach each other near
saddle points of the potential energy landscape. While classically the electron
cannot move from one cluster to another,  quantum mechanically it can
 tunnel through the potential barrier. If the electron energy $E$ is close 
 enough to the transition,  the
  potential barrier is close to parabolic and the tunneling probability through
  such as saddle point is proportional to ${\rm Exp}[-(E_c-E)]$. The number of
 such saddle points through which tunneling occurs
 in a system of length $L$ is typically $L/\ksi_p$.
Since the transmission coefficient is multiplicative, 
  the conductance (or the tunneling probability) through the whole system is
\begin{equation}
\sigma_{2D} \sim \left[ e^{-(E_c-E)} \right]^{L/\ksi_p} \equiv e^{-L/\ksi_{2D}} , 
\label{sigma2d}
\end{equation} 
with $\ksi_{2D}\sim (E_c-E)^{-\nu_{2D}}$ 
 and $\nu_{2D}= \nu_p + 1 = 7/3$. 
The best numerical estimate of the critical
exponent $\nu_{2D}=2.35\pm0.02$ \cite{huckenstein}, which is supported by
 experimental data \cite{wei}, 
  has a surprisingly excellent agreement with the result of the 
above argument, especially  in view of the crudeness of the argument.
 
 This argument can be generalized to the present problem \cite{phase},  as it
  is also expressed in terms of a two-dimensional Hamiltonian. In the presence
  of inter-layer tunneling (random field),  the only difference between the 
  present problem and the two-dimensional problem is the fact that the critical
  energy $E_c$ is not equal to the potential energy of the saddle-point,  but
  is $H_R$ away from it. Thus
\begin{equation}
\sigma_{3D} \sim \left[ e^{-H_R} \right]^{L/\ksi_p} \equiv e^{-L/\ksi_{3D}} , 
\label{sigma3d}
\end{equation} 
with $\ksi_{3D}\sim (E_c-E)^{-\nu}$ 
 and $\nu= \nu_p  = 4/3$. 
One finds the surprising result that the 
 critical exponent for the quantum three-dimensional problem is equal to the
 two-dimensional classical percolation exponent.
 This result is in excellent 
 agreement with existing numerical estimates,  $\nu=1.35\pm0.15$,  
 obtained both for a layered system and
  three-dimensional tight-binding model \cite{kramer},  and $\nu=1.45\pm0.25$, 
  obtained \cite{chalker} from a layered network model \cite{network}.

 Consider now the Hall conductance $\sigma_{xy}$. If the
 inter-layer tunneling $t$ is equal to zero, the system is a collection
of $N$ independent two-dimensional layers, all  with the same critical energy.
Thus $\sigma_{xy}$ will jump by
 $e^2/h$ in all layers simultaneously (see Fig. 4),  i.e., 
  it will have a single
  step of height $N e^2/h$ (which corresponds to a conductance per layer or
  conductivity of $e^2/h$). 
For finite $t$ (or finite random field) the situation is quite different.
 To see this we first carry out  a local $SU(N)$ gauge 
transformation in spin space, to rotate the spin by a unitary
matrix $U(x,y)$,
such that the $z$-direction always lies in the direction of the random
field. This exact transformation maps the Hamiltonian (\ref{hhamiltonian})
onto the equivalent Hamiltonian \cite{dkklee}
\begin{equation}
{\cal H} = ({\bf p}- e {\bf A}/c - i\hbar U^\dagger \nabla U)^2/2m + U(x, y) + 
{1\over S} S_z |H(x, y)|
\label{adiabatic}
\end{equation}
 If the potential energy and the inter-layer tunneling vary slowly in space, 
  one may apply the adiabatic approximation \cite{arovas}. In this approximation
  one neglects the additional $U^\dagger \nabla U$ term in the parentheses, 
  and  the Hamiltonian can be trivially diagonalized in spin-space. The
 random field serves as  an additional potential energy, which
  is different for each spin-direction (and its average is proportional
  to $H_R S_z$). Consequently, in this approximation
    one expects $N$ separate transitions,  each of the two-dimensional type 
 (see Fig. 4). 
 (Note that these transitions are not related to the different layers, 
 but rather
 to different coherent superpositions of the wave-functions in different layers).
 Since the separate transitions can only be resolved for energies
 smaller than $\Delta/N$, 
one expect  in this case a crossover from  a three-dimensional critical
 behavior, 
 for $|E-E_c|>\Delta/N$ to a two-dimensional critical behavior for 
 $|E-E_c|<\Delta/N$  (the two-dimensional behavior can only be seen for 
 temperatures smaller than $\Delta/N$),
\begin{equation}
\ksi = A_1\ \e^{-\nu_{2D}}f(\e) ,
\qquad f(\e)\rightarrow
\begin{cases}
1 & {\e\ll1} \\
{\displaystyle {A_2\ \e^{\nu_{2D}-\nu_{3D}}}
 } & {\displaystyle \e\gg1} , 
\end{cases}
\label{crossover}
\end{equation}
with $\e\equiv (E-E_c)/(\Delta/N)$. 
Thus, the effective exponent $\nu$ will 
crossover from its three-dimensional ($\sim4/3$) to the two-dimensional
value ($\sim7/3$), as one gets closer to the critical point from the insulating
side. Interestingly, for the case $\nu_{2D}=7/3$ and $\nu_{3D}=4/3$ the scaling
function $f(\e)$ may be analytic. This crossover can be studied via the critical
behavior of the conductance (Eqs.(\ref{sigma2d}) and (\ref{sigma3d})),  or
by that of $d \sigma_{xy}/dB$ \cite{simon}. 

\vskip 1.0 truecm
\begin{center}
\leavevmode
\epsfxsize=3.3in
\epsfbox{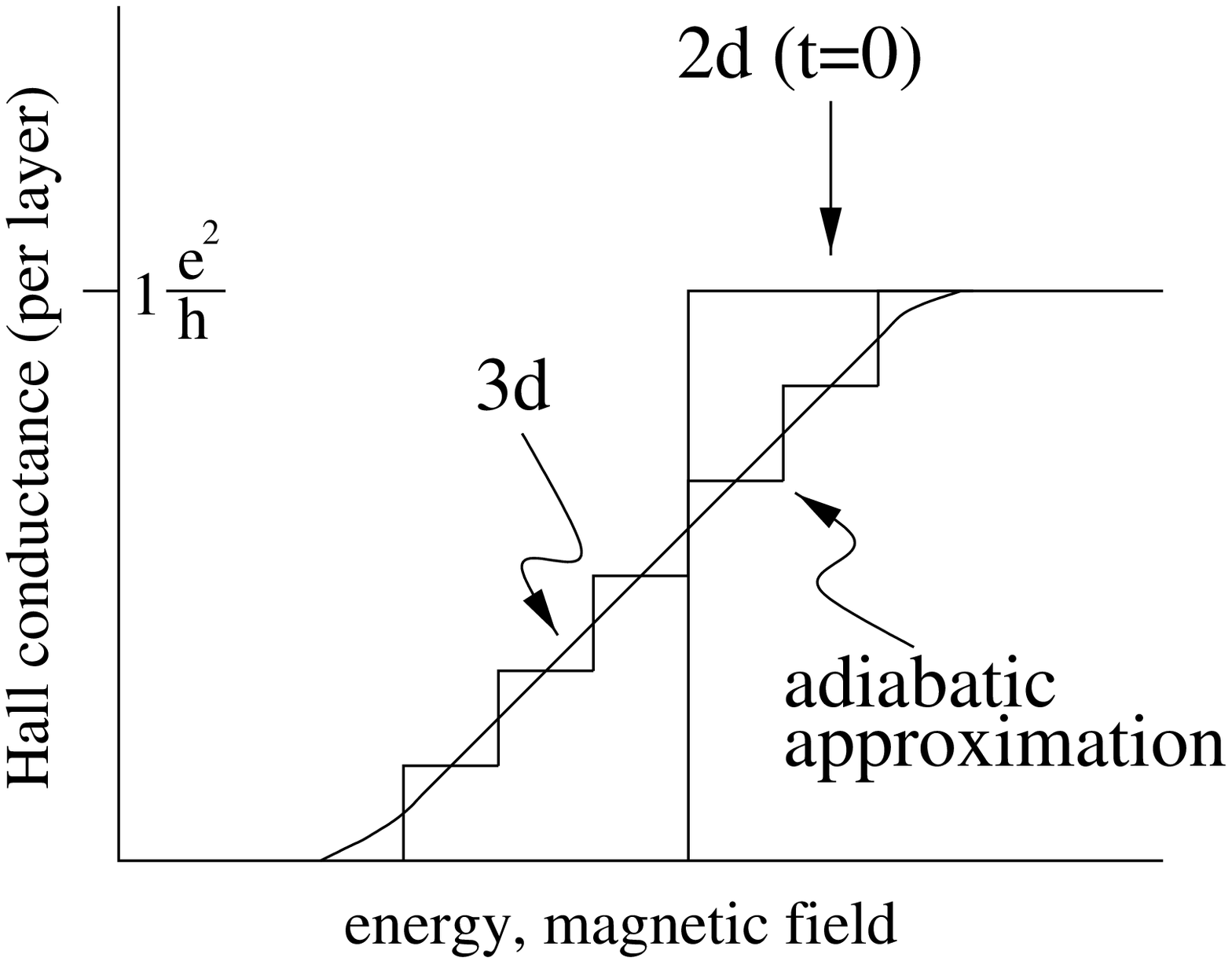}
\end{center}
\begin{small}
Fig. 4. The change in the Hall conductance at the transition. For $t=0$ there is
 a two-dimensional behavior (a single step). For $t\ne0$,  then in the adiabatic
 approximation one find a series of smaller steps,  of the number of layers 
 (see text). In the three dimensional limit, as the number of layers increases,  
 one expects a smooth transition between the quantized values (a metallic region).
\end{small}
\vskip 0.5 truecm

 In the adiabatic approximation there is a zero-temperature 
 metallic phase only in the true
 three-dimensional limit ($N\rightarrow\infty$),  which is the classical limit
 ($S\rightarrow\infty$) of the spin-problem (Fig. 4). As non-adiabaticity
(the additional term in the parentheses in Eq.(\ref{adiabatic})) is switched on,
 the different
spin-states that were the eigenstates of the system in the adiabatic limit
get coupled. It is not clear if this coupling will smear out the separate 
transitions even for a finite number of layers. 
It is known
  that there may 
  occur transitions between the expected adiabatic behavior to a different
 behavior (as a function of e.g. the tunneling matrix element),  even for the 
 two-layer problem \cite{2layer},  and it remains to be seen if such a deviation
 from the adiabatic limit will  also occur for a finite number of layers. We hope
 that this work will motivate further studies in this direction.
 
To conclude, we have used a mapping onto a two-dimensional
 spin-Hamiltonian to describe the
physics of the quantum Hall effect in three-dimensional layered systems. This
mapping was used mainly for conceptual reasons, in order to allow us to
extend methods applied in the traditional two-dimensional quantum Hall systems to
 the present case. The arguments presented here, however, could be directly
applied to the original three-dimensional system, and thus none of the results
of this paper depends on the particular form of the spin-Hamiltonian. For
example, in the three-dimensional layered system, 
the potential and the hopping part
of the Hamiltonian (Eq.(\ref{hamilonian})) can be recast in a form of 
a position-dependent $N\times N$ matrix. Diagonalizing this matrix locally and
carrying out a unitary local rotation in layer-space,  will lead to a Hamiltonian
of the form (\ref{adiabatic}),  and to all the results of the last section.
Similar arguments can be made to derive the phase-diagram and the critical 
exponent.
 
The author thanks A. Stern for several discussions.

\end{multicols}

\begin{references}
\bibitem{but}
Note,  however, that the Hamiltonian (\ref{spinhamilonian}) has less degrees
of freedom than Hamiltonian (\ref{hamilonian}),  since at each point in the
plane the potential in spin-space forms a ladder,  spaced by energy 
$\Delta U(x, y)$. Nevertheless,  if one imagine coarse-graining this Hamiltonian, 
then at each iteration more and more frequencies will appear and instead of
 a equally-spaced ladder,  one will have now quasi-periodicity,  with increasing
 number of frequencies. Thus one may expect that under such renormalization, 
 the Hamiltonian (\ref{spinhamilonian}) will flow to the same fixed point as
 (\ref{hamilonian}), a fact which is borne out by the phase diagram and the
 critical exponent which seem to be identical. Similarly, the weak dependence 
of the tunneling on the layer number is expected not to be a relevant
perturbation.
\bibitem{exponent}
In fact, the way the width of the metallic region goes
 to zero is related to the critical behavior
of the two-dimensional localization length,  see Ref.\cite{chalker}.
\end{references}
\end{document}